\documentclass[preprint,a4paper,3p,10pt,times,twocolumn,dvipdfm]{elsarticle}
\usepackage{graphics}
\usepackage{graphicx}
\usepackage{epsfig}
\usepackage{amssymb}
\usepackage{amsthm}
\usepackage{natbib}
\usepackage{latexsym}
\usepackage{mathrsfs}
\usepackage{amsmath}
\usepackage{amscd}
\usepackage{color}
\usepackage{verbatim}
\biboptions{sort&compress}
\journal{Physics Letters A}
\begin{document}
\begin{frontmatter}
\title{Bifurcation structures and transient chaos in a four-dimensional Chua model}
\author{Anderson Hoff}
\ead{hoffande@gmail.com}
\author{Denilson T. da Silva}
\author{Cesar Manchein}
\ead{cesar.manchein@udesc.br}
\author{Holokx A. Albuquerque$^*$}
%
\address{Departamento de F\'\i sica, Universidade do Estado de Santa Catarina, 89219-710 Joinville, Brazil}
%
%
\begin{abstract}
  A four-dimensional four-parameter Chua model with cubic nonlinearity is studied
  applying numerical continuation and numerical solutions methods. Regarding
  numerical solution methods, its dynamics is characterized on Lyapunov and isoperiodic
  diagrams and regarding numerical continuation method, the bifurcation curves
  are obtained. Combining both methods the bifurcation structures of the model
  were obtained with the possibility to describe the {\it shrimp}-shaped domains
  and their endoskeletons. We study the effect of a parameter that controls the
  dimension of the system leading the model to present transient chaos with its
  corresponding basin of attraction being riddled.
\end{abstract}
%
\begin{keyword}
Four-dimensional model \sep Bifurcation curves \sep Transient chaos \sep Shrimps.\\
$^*$Corresponding author. Tel.: +55 47 4009 7721; fax: +55 47 4009 7940. Email: holokx.albuquerque@udesc.br
\end{keyword}
\end{frontmatter}
%
%
\section{Introduction}
\label{introduction}

In dynamical systems a bifurcation structure is a change of qualitative behavior
when control parameters are varied. Recently, a plethora of bifurcation structures
is being reported in many types of discrete- and continuous-time nonlinear
dynamical systems, when two or more control parameters are varied
\cite{alb,alan,cris,stoop,denis,den,med,mede,gal1,wil,gal5,manc2,manc,manc1,rech3}.
In such works and references therein a common feature
observed in the parameter-planes is the occurrence of isoperiodic stable structures,
often namely {\it shrimp}-shaped domains (SSD's), with a variety of bifurcation cascades
and organization rules. In the last years, many works \cite{wil,bar1,bar2,ray,gen}
use numerical continuation methods to unveil in more details the bifurcation
structures in the parameter-planes of low order dissipative systems.
In particular, the authors of a recent Letter \cite{wil} use numerical continuation techniques
to study in details a periodic domain in the biparametric-space of the
two-dimensional H\'enon map. In that work, the authors claimed that
using continuation methods is more feasible to analyze the delicate bifurcation
structures for both nonlinear differential equations and maps, than use shooting
methods.

Regarding the Chua circuit model \cite{alb,cris,gal8,mar,chua}, a paradigm of
chaotic behavior, in the last four decades an enormous quantity
of published papers report the rich dynamics that this circuit presents.
The canonical Chua model consists of three coupled first-order ordinary differential
equations with a piecewise linear function. Other smooth nonlinear functions are
also used \cite{gal8,alb1,cris1}. In all these three-dimensional Chua models
the bifurcation structures in the parameter-planes present different types
of roles and organizations. Based on the canonical three-dimensional Chua model,
high-dimensional versions can be constructed introducing feedback controllers
\cite{cris,liu}, {\it i.e.}, adding another variables in the model. In this
configuration, the high-dimensional model can present more complex dynamics,
as hyperchaos \cite{cris}.

Our main goal in this Letter is to apply numerical continuation methods \cite{kuz}
in addition with numerical solution methods to unveil the bifurcation
structures in two-dimensional parameter-spaces, and the endoskeletons of SSD's,
in a high-dimensional continuous-time nonlinear model.
In this work we show that the endoskeleton of a SSD is
formed by four main bifurcation curves: two saddle-node curves with one cusp point
and two intersected period-doubling curves. These four curves delimit the
lowest-period of the structure. Another remarkable observation is the presence
of transient chaos in the two-dimensional parameter-spaces of the model.
As far as we know, the work here reported is one of the first to combine both
numerical continuation and numerical solution methods in high-dimensional
continuous-time nonlinear model to discuss the bifurcation structures
presented by the model.

This paper is organized as follows: In Section \ref{mod} we present the model
with a brief description of the employed methods. In Section \ref{res}
we present the numerical results with some discussions, and in
Section \ref{conc} we present the conclusions of this work.

\section{The four-dimensional model}
\label{mod}

The nonlinear model studied has the origin in the canonical
three-dimensional three-parameters Chua model with cubic nonlinearity,
adding a controller variable $w$ and a fourth parameter $d$ as
proposed in a recent paper \cite{liu}. The nonlinear equations are
\begin{equation}\label{chua4d}
\begin{split}
\dot x &= \frac{dx}{dt} = a(y + 0.2(x - x^3)), \\
\dot y &= \frac{dy}{dt} = bx - y + z + w, \\
\dot z &= \frac{dz}{dt} = -cy + w, \\
\dot w &= \frac{dw}{dt} = -dy,
\end{split}
\end{equation}
where $x$, $y$, $z$, and $w$ are the variables, that represent the
voltages in the real circuit \cite{liu}, $a$, $b$, $c$, and $d$
are the parameters, related to combinations of resistors and
capacitors in the real circuit. Such system is a
four-parameter model and the parameter $d$ controls the extra-dimension,
once that if $d$ vanish, the last differential
equation of the system~(\ref{chua4d}) vanish, too. Therefore,
if we initialize the system~(\ref{chua4d}) with the initial
conditions $(x, y, z, w) = (0.1,0.1,0.1,0.0)$, we will
have the canonical three-dimensional Chua system, as reported in the
Refs.~\cite{alb,gal8,mar,chua,cris1}.

It is well known that for a four-dimensional system, the Lyapunov
spectrum has four values of exponent, $(\lambda_1, \lambda_2, \lambda_3, \lambda_4)$,
one value for each direction of the flow. Here, we use the largest
exponent of the spectrum to study in the parameter-spaces
(henceforth Lyapunov diagrams) the behaviors
of the system~(\ref{chua4d}) identifying colors to the
exponent values. For this purpose, the four-dimensional
Lyapunov diagram, $a \times b \times c \times d$, is sliced
in the following parameter combinations: $a \times b$ with $c$
and $d$ fixed, $a \times c$ with $b$ and $d$ fixed,
$b \times c$ with $a$ and $d$ fixed, $a \times d$ with $b$ and $c$ fixed,
$b \times d$ with $a$ and $c$ fixed, and $c \times d$ with $a$ and $b$ fixed.
To obtain the Lyapunov spectra, the model is numerically
solved by the Runge-Kutta method with fixed time step equal
to $1.0 \times 10^{-2}$ and $5.0 \times 10^5$ integrated steps to obtain
the exponent spectrum via the algorithm proposed in \cite{wolf}
for each parameter pair discretized in a grid of $500 \times 500$
values. Therefore, we obtain $2.5 \times 10^5$ Lyapunov exponent
spectra for each parameter pair.

On the other hand, the bifurcation curves are obtained by continuation methods,
using a properly continuation package for continuous-time systems,
as MATCONT \cite{kuz}, which is a standard tool for numerical
bifurcation analysis. Here we superimpose the bifurcation
curves of Hopf, saddle-node, and period-doubling on the
Lyapunov diagrams to unveil some bifurcation structures
presented by the system~(\ref{chua4d}) in the parameter-space.

The parameter-space of periods (henceforth isoperiodic diagrams),
is used to corroborate the period-doubling and saddle-node
bifurcation structures presented by the system~(\ref{chua4d}).
The periods are obtained by the maxima of the time-series
for a given set of parameters, as done in Ref.~\cite{gal1}.
To evaluate the period we use the Runge-Kutta method with
variable time step, removing a transient time equal
to $5 \times 10^6$ and using more $1 \times 10^6$ iterations to find
the period with a precision of $1 \times 10^{-3}$.
The parameter pairs were discretized in a grid of
$1200 \times 1200$ values.

\section{Results}
\label{res}

\begin{figure}[htb]
  \centering
  \includegraphics*[width=1.00\columnwidth]{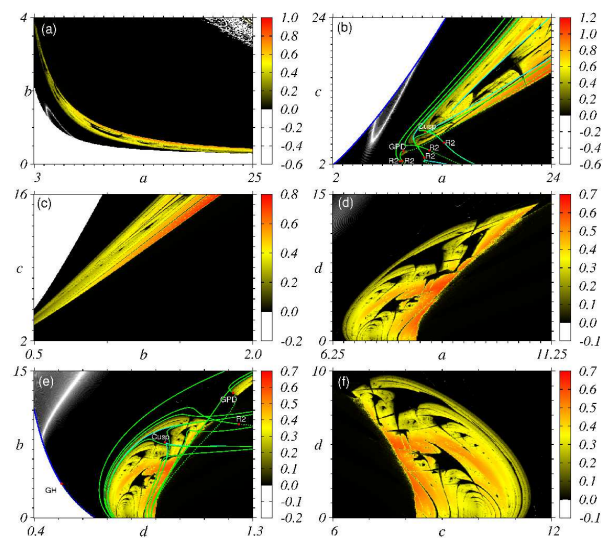}
  \caption{(Color online) Lyapunov diagrams for the largest exponent of the Lyapunov
  spectrum codified by colors, as the color bars at the right side
  of each diagram, for the six combinations of parameters of the
  system~(\ref{chua4d}). (a) $(a \times b)$ with $c = 12.5$
  and $d = 5.0$, (b) $(a \times c)$ with $b = 0.55$ and $d = 10.0$,
  (c) $(b \times c)$ with $a = 7.0$ and $d = 1.0$,
  (d) $(a \times d)$ with $b = 0.88$ and $c = 8.5$,
  (e) $(b \times d)$ with $a = 8.0$ and $c = 8.5$, and
  (f) $(c \times d)$ with $a = 8.0$ and $b = 0.88$. The blue,
  green, and cyan curves in (b) and (e) are Hopf, period-doubling,
  and saddle-node bifurcation curves, respectively.}
  \label{fig1}
\end{figure}

Figure~\ref{fig1} shows the Lyapunov diagrams for six slices of the
four-dimensional parameter-space of the system~(\ref{chua4d}).
In the diagrams the largest value of the Lyapunov spectrum is
codified in colors following the color bars at the right side.
White, black and yellow/red regions in the diagrams refer to the
equilibrium points, periodic points, and chaotic behaviors of
system~(\ref{chua4d}), respectively. In these diagrams, an abundant
presence of periodic structures (black regions) immersed in chaotic
regions (yellow/red regions) can be observed, as well as large regions
of equilibrium points (white regions). Some periodic structures observed in
Fig.~\ref{fig1} are reported in others works for different types of systems,
indicating that these structures are general, moreover, these
structures are organized in specific directions on the Lyapunov diagrams.
Some of these structures are known by SSD's \cite{gal9}.
In (b) and (e) the Hopf, period-doubling, and saddle-node curves,
blue, green, and cyan lines, respectively, are superimposed.
The Hopf curve delimits the equilibrium points (white region) to limit cycles
(black region) bifurcations, the period-doubling curves delimit
the regions where occur a doubling of the limit cycles, and the
saddle-node curve delimits the points where a chaotic attractor
disappears and a stable periodic attractor is created. The solid
and dashed green lines are stable and unstable period-doubling curves,
respectively. The red points R2 and GPD are: 1:2 resonance point
where two eigenvalues are equal to $-1$, and generalized period-doubling
point where one eigenvalue is equal to $-1$ and the normal form
coefficient is null, respectively, and denote the location of
stability loss of the period-doubling curves. The bifurcation
curves shown in Figs.~\ref{fig1}(b) and~\ref{fig1}(e) reveal the
bifurcation structures in a two-dimensional parameter-space and
as can be seen in such figures, the curves delimit the boundaries
between the stable points and chaotic domains, as well as the
endoskeleton of the SSD's. The bifurcation curves for the other Lyapunov diagrams in
Fig.~\ref{fig1} were omitted by sake of simplicity, because
the bifurcation structures in these diagrams not present new
information of those presented in the exemplary cases of
Figs.~\ref{fig1}(b) and~\ref{fig1}(e).

\begin{figure}[htb]
  \centering
  \includegraphics*[width=1.00\columnwidth]{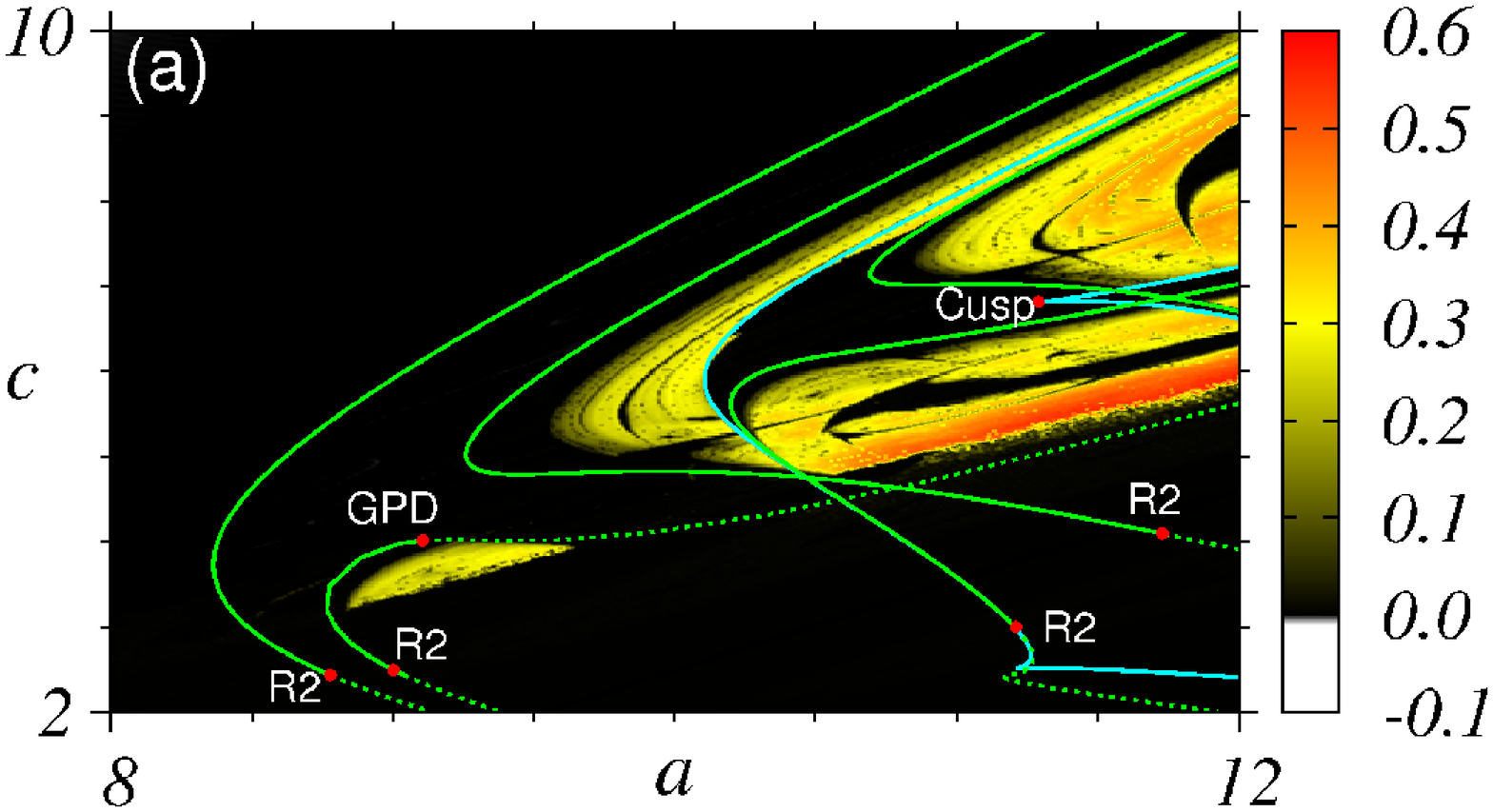}
  \includegraphics*[width=1.00\columnwidth]{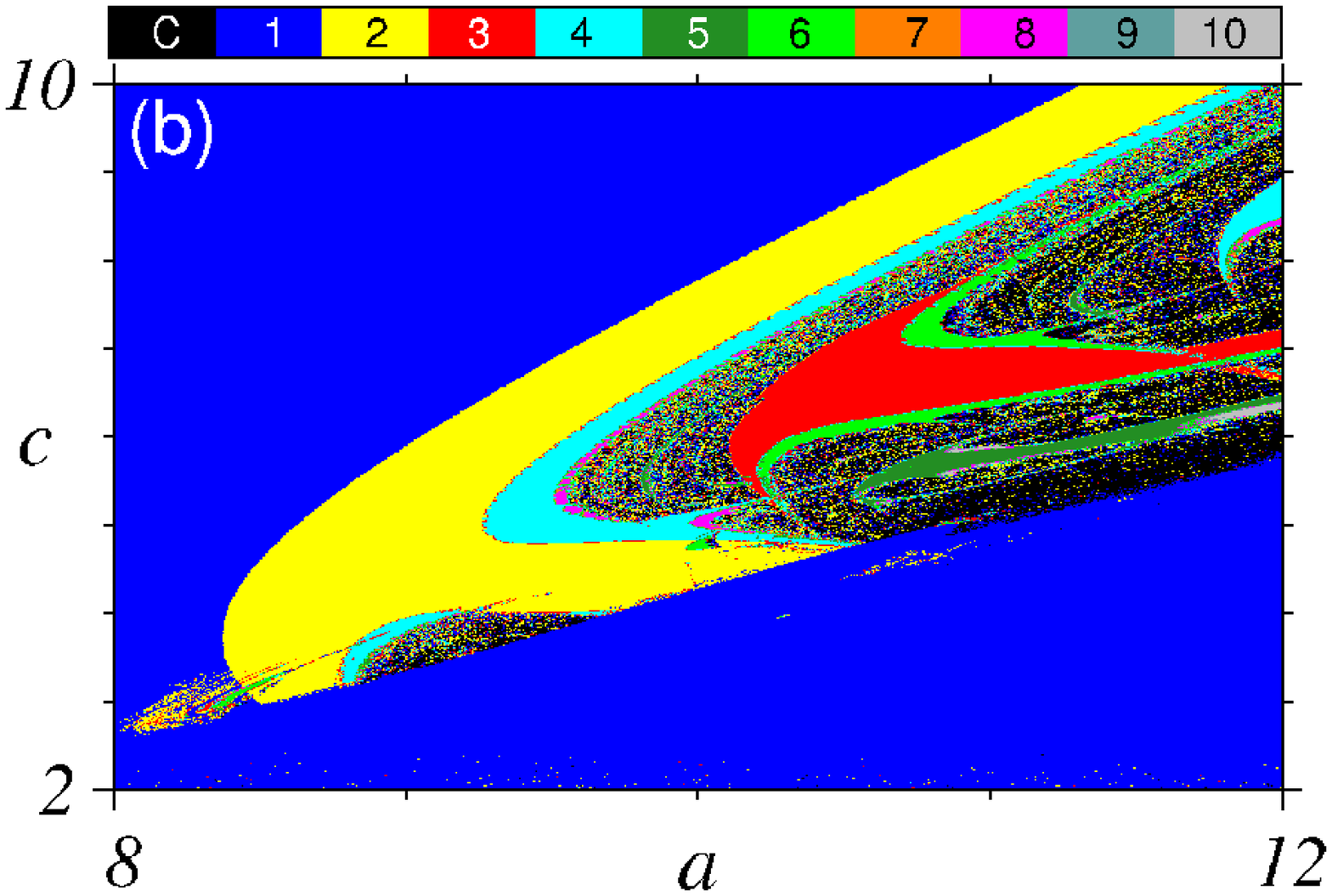}
  \caption{(Color online) In (a) Lyapunov diagram for the $a \times c$ plane with
  $b = 0.55$ and $d = 10.0$, for the largest Lyapunov exponent with
  bifurcation curves superimposed, and in (b) isoperiodic diagram for
  the same ranges in (a). The top color palette in (b) codifies the
  periods. Black color codifies chaotic behavior or periods greater
  than 10.}
  \label{fig2}
\end{figure}

\begin{figure}[htb]
  \centering
  \includegraphics*[width=1.00\columnwidth]{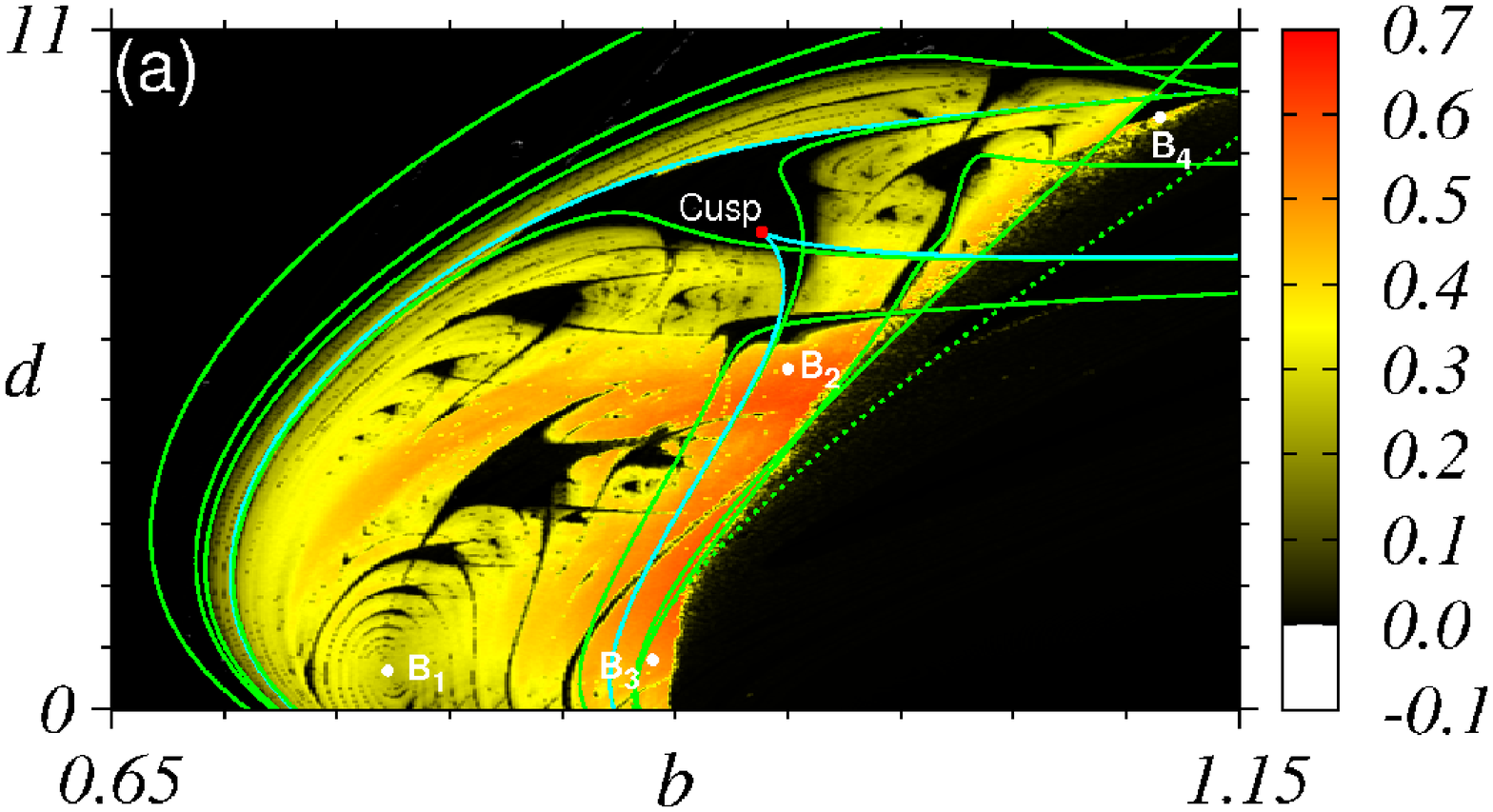}
  \includegraphics*[width=1.00\columnwidth]{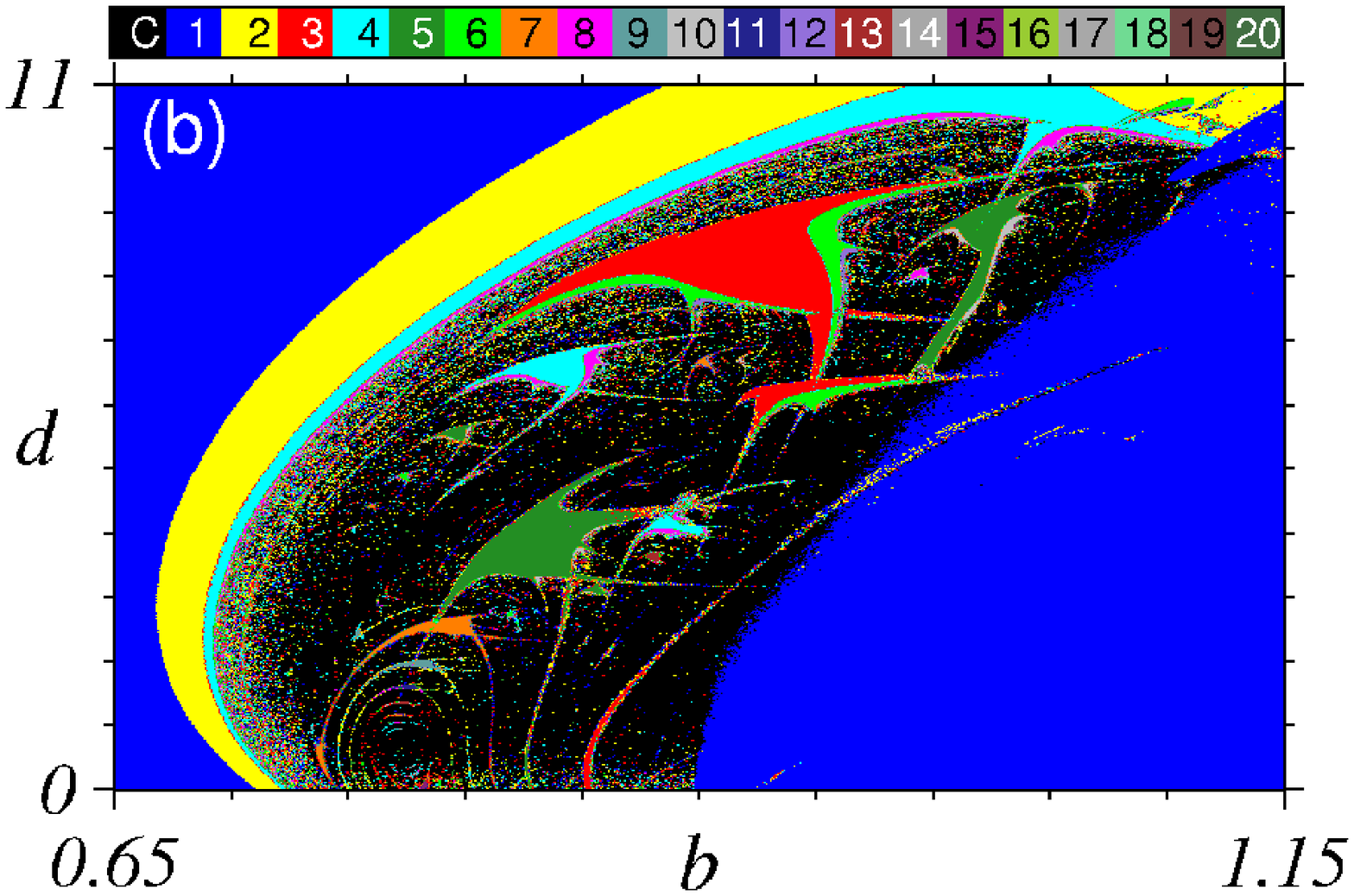}
  \caption{(Color online) In (a) Lyapunov diagram for the $b \times d$ with $a = 8.0$
  and $c = 8.5$, for the largest Lyapunov exponent with bifurcation
  curves superimposed, and in (b) isoperiodic diagram for the same
  ranges in (a). The points B$_1, \cdots, $B$_4$ in (a) will be discussed
  in Figs.~\ref{fig8} and~\ref{fig9}. The top color palette in (b)
  codifies the periods, in which black color codifies chaotic behavior
  or periods greater than 20.}
  \label{fig3}
\end{figure}

To present, in more details, the bifurcation structures of the
system~(\ref{chua4d}), we show in Figs.~\ref{fig2} and~\ref{fig3} the Lyapunov
and isoperiodic diagrams for the $a \times c$ plane with $b = 0.55$ and
$d = 10.0$, and for $b \times d$ plane with $a = 8.0$ and $c = 8.5$, respectively.
These diagrams are amplifications of Figs.~\ref{fig1}(b) and~\ref{fig1}(e).
The period-doubling and saddle-node bifurcation curves, in green and cyan,
respectively, obtained by numerical continuation method, are superimposed
on the Lyapunov diagrams and reveal the set of points in the parameter-space
where period-doublings and sudden changes of chaotic to periodic motion occur.
The isoperiodic diagrams, obtained by numerical solution method, corroborate and
identify the position of these bifurcation curves, and both reveal the bifurcation
structures by period-doubling and by saddle-node in two-dimensional
parameter-spaces of the system~(\ref{chua4d}).

In Fig.~\ref{fig2}(a) we present in more details the period-doubling
bifurcation curves, green curves, and the endoskeleton of the
period-$3$ SSD (see the red structure in Fig.~\ref{fig2}(b)),
{\it i.e.}, two saddle-node bifurcation curves (cyan curves) with
the cusp point in addition with two intersected period-doubling
curves (green curves). It is clear to see that these curves delimit
regions in the parameter-space that have same periods, as can be observed
in Fig.~\ref{fig2}(b) (yellow and red domains). In these cases,
the curves delimit the domain of lowest period of the SSD's,
that is inlaid in the period-$1$ domain (see the blue region in
Fig.~\ref{fig2}(b)) and that is immersed in the chaotic domain
(see the red structure in Fig.~\ref{fig2}(b)). In the Lyapunov diagram of
Fig.~\ref{fig2}(a) without the bifurcation curves, it is impossible
to identify the ingrown SSD of period-$2$.

Regarding Fig.~\ref{fig3}, an amplification of Fig.~\ref{fig1}(e),
the same features observed in Fig.~\ref{fig2} is also observed
with additional remarks. For instance, in (a) we observe two
solid green curves crossing the chaotic domain at right and
bordering two distinct periodic structures, the structures
of period-3 and 5 in (b), near the period-$1$ border at right.
Inside these structures, the green bifurcation curves delimit
the points of period-doubling bifurcation. Indeed, for the
period-$3$ structure (red structure in Fig.~\ref{fig3}(b)),
the period-doubling bifurcation occurs near the lateral border,
corroborating the result shown in Fig.~\ref{fig3}(a). The same
occurs for the period-$5$ structure (dark green structure in
Fig.~\ref{fig3}(b)). The endoskeleton of the bigger period-$3$
structure in Fig.~\ref{fig3}(b) is also clearly visible
in Fig.~\ref{fig3}(a) with the cusp point inside it.

\begin{figure}[htb]
  \centering
  \includegraphics*[width=1.00\columnwidth]{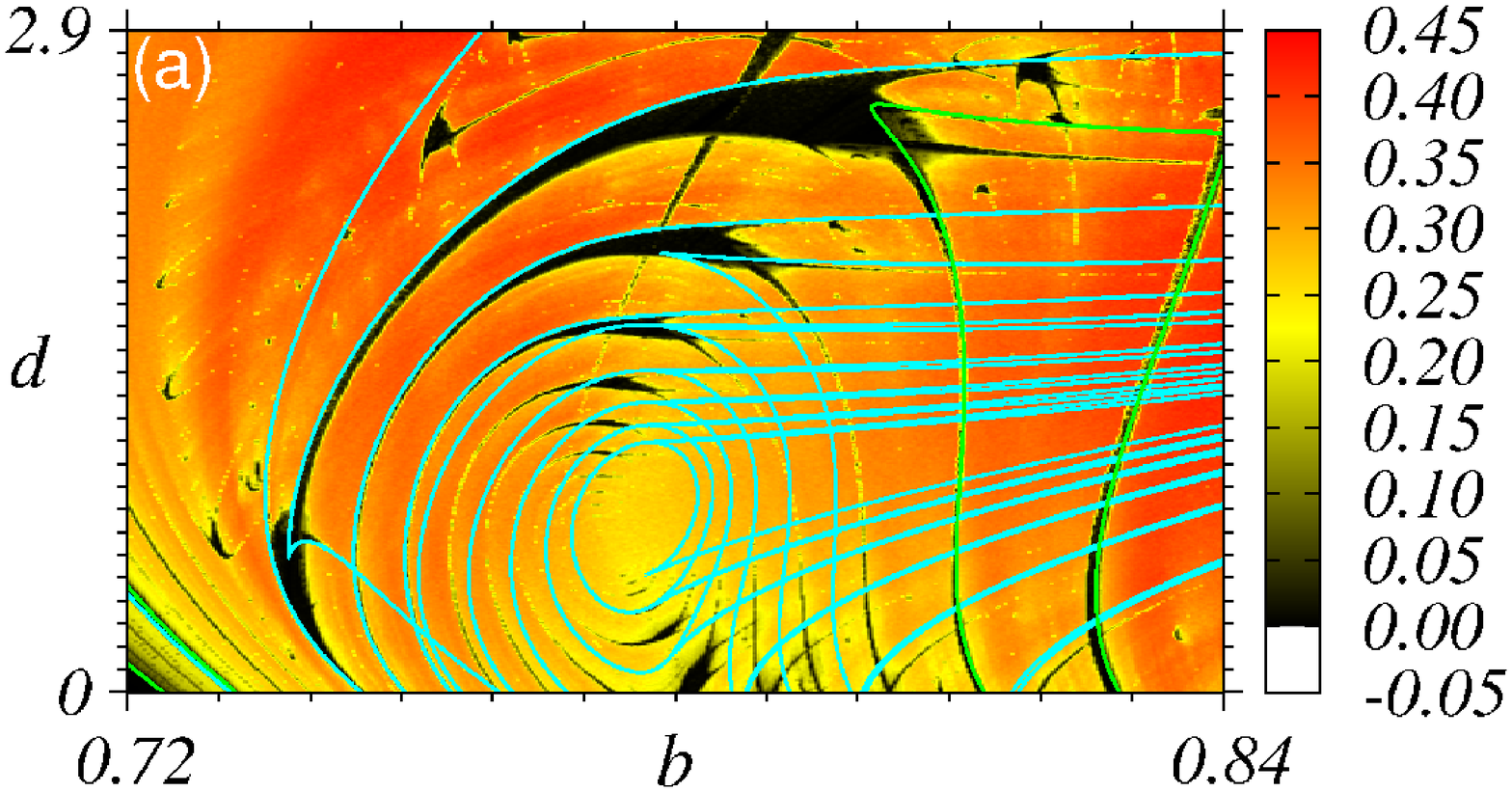}
  \includegraphics*[width=1.00\columnwidth]{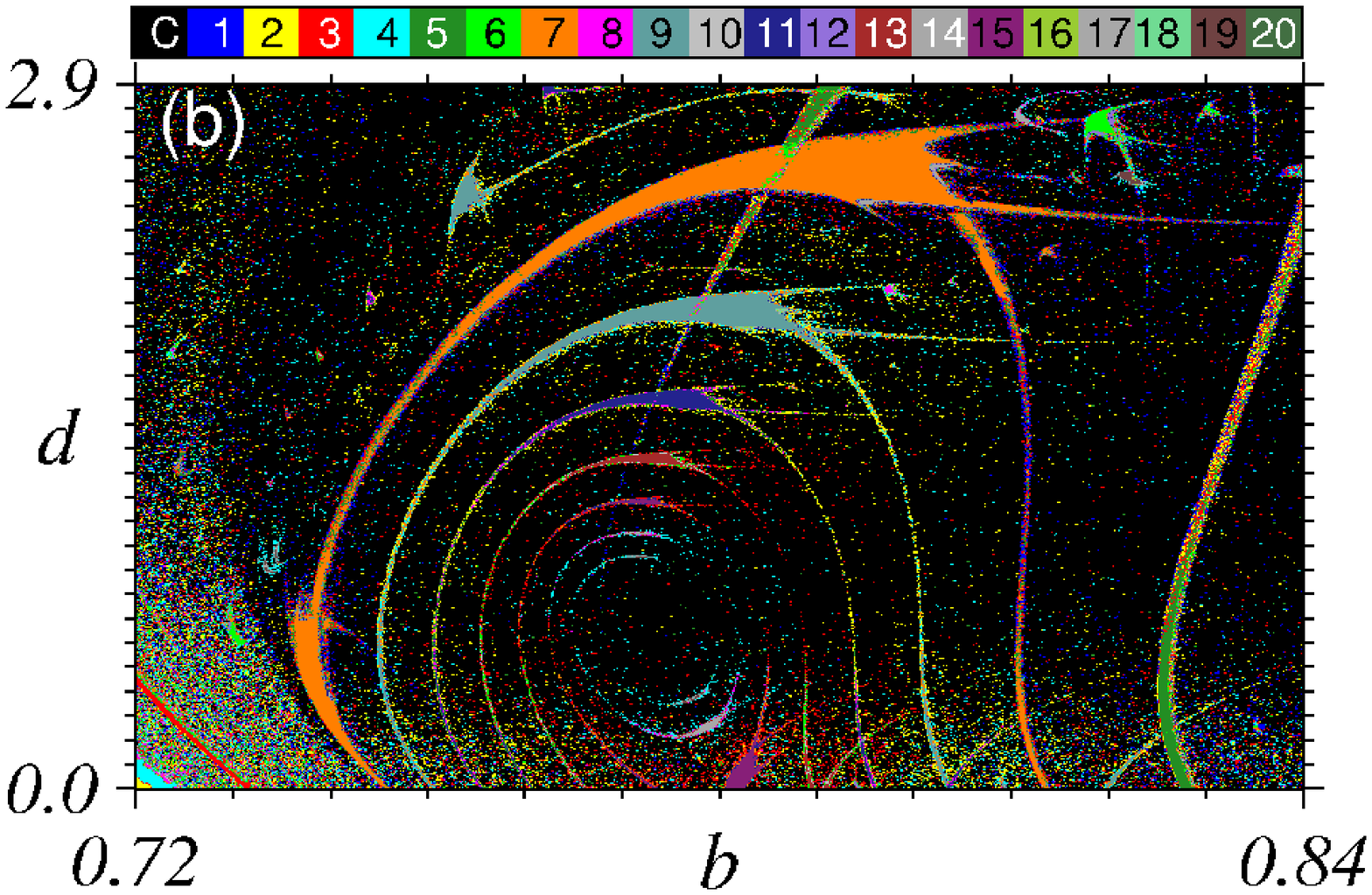}
  \caption{(Color online) Amplification of Fig.~\ref{fig3} emphasizing the periodicity
  spiral of SSD's. In (a) the Lyapunov diagram,
  and (b) the isoperiodic diagram, in which black color codifies chaotic
  behavior or periods greater than 20. Some bifurcation curves are
  superimposed in the Lyapunov diagram in (a). Green for period-doubling
  and cyan for saddle-node bifurcations.}
  \label{fig4}
\end{figure}

Other organization structure observed in a two-dimensional
parameter-spaces of system~(\ref{chua4d}), is the presence
of a periodicity spiral, {\it i.e.}, self-connected
periodic structures that coil up around a focal point.
This behavior is shown in Fig.~\ref{fig4}, where it is
an amplification of Fig.~\ref{fig3}. The periodicity spiral
was reported in some recent works about three-dimensional Chua's model \cite{gal8,alb2},
in other systems \cite{slip,gal10}, and references therein.
Here, we show the periodicity spiral for a four-dimensional
Chua's model, and the dimensional parameter $d$, that controls
the additional dimension of system~(\ref{chua4d}), has a fundamental role
in this organization. This organization structure is
observed only in parameter-spaces in which the parameter
$d$ is varying, as can be observed in Figs.~\ref{fig1}(d)-(f).
In Fig.~\ref{fig4}(b), we also observe the organization
of the periods that increase its value $(7 \to 9 \to 11 \to 13 \cdots)$ as the structures
coil up around a focal point, therefore the bifurcation
structure of this spiral follows a period-adding cascade.
Bifurcation curves are also superimposed in Fig.~\ref{fig4}(a),
with period-doubling curves in green and saddle-node curves in cyan.
The skeleton of this spiral structure is the connected
period-doubling and saddle-node curves. These bifurcation
structures of the spiral follow the same structure reported
earlier for three-dimensional continuous-time models \cite{bar2,gal8,gal9,alb2,slip}
and maps \cite{gasp}.

Previous works \cite{cris,cris1} report some Lyapunov diagrams
for a three- and four-dimensional version of Chua system,
both with cubic nonlinearity. In Ref.~\cite{cris1}, the
three-dimensional Chua system presented a well-known dynamics,
with SSD's embedded in all two-dimensional
Lyapunov diagrams, self-organizing by period-adding bifurcation
cascades in specific directions in the parameter-spaces, and
accumulate in periodic borders. A remarkable difference emerged
in the hyperchaotic four-dimensional Chua system reported in
Ref.~\cite{cris}, where {\it malformed}-SSD's
become present. This malformation is explained in Ref.~\cite{barr} 
using concepts borrowed from the Windows Conjecture \cite{bar}, 
that establishes the relation
between the branches of the domains near hyperchaotic regions.
In few words, the conjecture says that in hyperchaotic systems
the branches of the SSD's are limited in
the parameter-spaces, in the sense that the branches are broken,
and in chaotic systems the branches are extended, {\it i.e.},
the SSD's extend all over the parameter-spaces.
For more details see Refs. \cite{cris,barr}. It is worth to
mention here, in the set of parameters studied for system~(\ref{chua4d})
and presented in this Letter, we do not observe hyperchaotic
behavior, in consequence, all the SSD's
present extended branches, as can be observed in Fig.~\ref{fig1}.
\begin{figure*}[htb]
  \centering
  \includegraphics*[width=2.00\columnwidth]{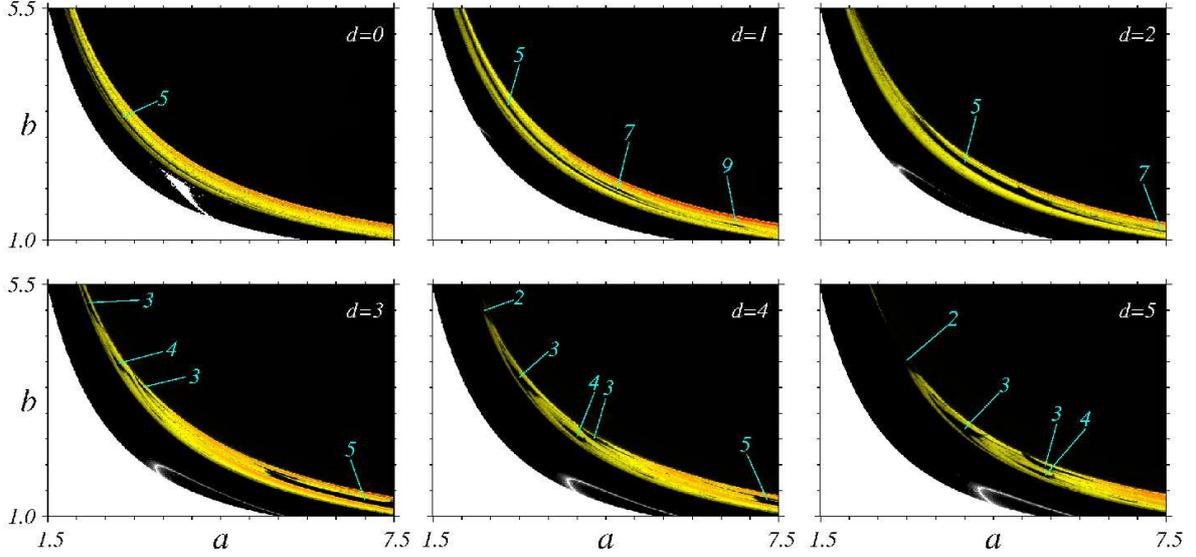}
  \caption{(Color online) Lyapunov diagrams for the $(a \times b)$ plane with $c = 12.5$,
  for the largest Lyapunov exponent with the parameter $d$, whose values
  are shown in each panel. The cyan numbers are the periods of each
  structure.}
  \label{fig5}
\end{figure*}
The system~(\ref{chua4d}) presents another interesting feature
regarding the dimensional control parameter $d$, that is responsible
by the coupling of the three-dimensional Chua model with an
extra-dimension. To discuss the influence of $d$, we show in
Fig.~\ref{fig5} the Lyapunov diagrams for
six values of $d$. For $d=0$, no SSD's is
visible on the chaotic domain but there exist just black stripes
that indicate periodic domains, however
as $d$ increases the structures appear and become bigger,
in addition they present a movement in the chaotic
domain as can be observed by their change of position
in the diagrams. The plots of Fig.~\ref{fig5} show how
the dynamics change with the dimensional control
parameter $d$ in the system~(\ref{chua4d}). A clear
evolution of the SSD's since their births, growths
and movements can be observed.

\begin{figure}[htb]
  \centering
  \includegraphics*[width=1.00\columnwidth]{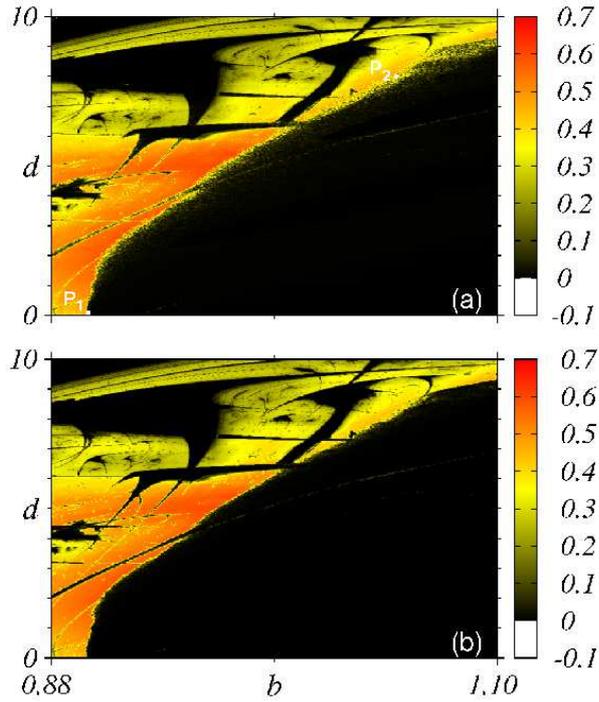}
  \caption{(Color online) Amplification of Fig.~\ref{fig3} emphasizing the right
  periodic border, in black, where transient chaos is visible.
  In (a) the transient time is $5.0 \times 10^5$, and in (b) is
  $5.0 \times 10^6$.}
  \label{fig6}
\end{figure}

\begin{figure}[htb]
  \centering
  \includegraphics*[width=1.00\columnwidth]{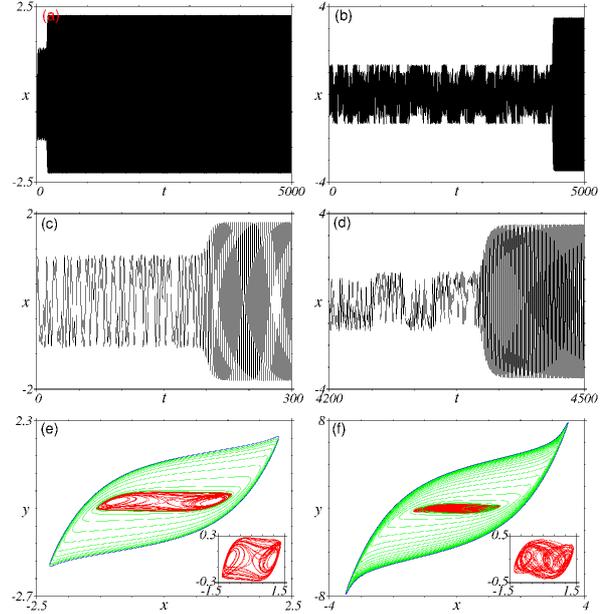}
  \caption{(Color online) Left column, (a),(c), and (e), transient chaos in the point
  P$_1$ of Fig.~\ref{fig6}(a). Right column, (b), (d), and (f), transient
  chaos in the point P$_2$ of Fig.~\ref{fig6}(a).
  In (a)-(d) the time series, and in (e)-(f) the chaotic and periodic
  attractors with the transient between them. The insets show in
  more details the chaotic attractors.}
  \label{fig7}
\end{figure}

Other interesting feature presented by the system~(\ref{chua4d}) is
transient chaos, characterized by topology changes in the attractors after
a transient time \cite{tel}, that is observed in the $b \times d$
parameter-space. To illustrate this behavior,
in Fig.~\ref{fig6} we show the Lyapunov diagrams emphasizing the right
period-$1$ border of Fig.~\ref{fig3}. In (a) a transient time of
$5.0 \times 10^5$ was used to evaluate the largest Lyapunov exponent,
and in (b) a transient time of $5.0 \times 10^6$ was used. It is clear to
observe the differences between the right black borders in both plots,
for example, for $d$ greater than zero in (a) the border is grained and
in (b) the border is sharp. Otherwise, for $d$ near zero in (a) and (b)
the borders are more sharp. To illustrate these features we show in
Fig.~\ref{fig7} the changes of chaotic to periodic behavior
for two points in Fig.~\ref{fig6}(a), namely P$_1$ and P$_2$ points,
respectively.

In the left column of Fig.~\ref{fig7}, (a), (c), and (e), we show the time
series for the parameters $(b,d) = (0.898,0.1)$ (see the point P$_1$ in Fig.~\ref{fig6}(a)).
In Fig.~\ref{fig7}(c) the time series is an amplification of Fig.~\ref{fig7}(a),
to show clearly the change of chaotic to periodic behavior.
In Fig.~\ref{fig7}(e) we present the behavior of the attractor,
beginning in chaotic oscillations (see the red lines) and shortly
passing to periodic oscillations (see the green lines).
In the right column of Fig.~\ref{fig7}, (b), (d), and (f), the time series is for
$(b,d) = (1.05,8.0)$ (see the point P$_2$ in Fig. 6(a)).
In Fig.~\ref{fig7}(d) the time series is an amplification of
Fig.~\ref{fig7}(b), to show the change between chaotic and periodic
behavior. In Fig.~\ref{fig7}(f) the attractors are shown. See in more
details the chaotic attractors in the insets of Figs.~\ref{fig7}(e) and~\ref{fig7}(f).
In this case, the chaotic oscillations last longer than the
previous case (compare the Figs.~\ref{fig7}(a) and~\ref{fig7}(b)).
Following the definitions of the book \cite{tel}, the two examples
presented follow the case in which the nonattracting chaotic
set coexists with an asymptotic periodic attractor.

\begin{figure}[htb]
  \centering
  \includegraphics*[width=1.00\columnwidth]{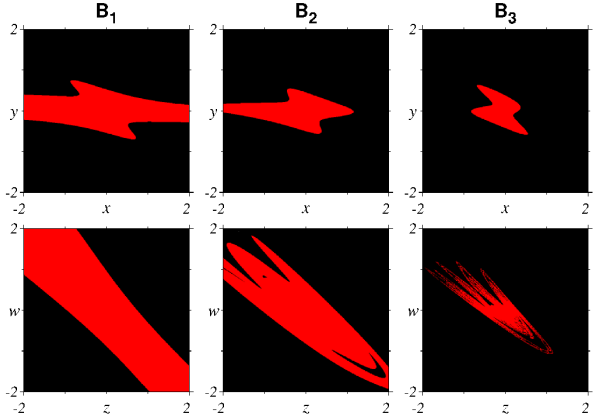}
  \caption{(Color online) Basins of attraction for the points B$_1$, B$_2$, and B$_3$ in
  Fig.~\ref{fig3}(a). Each column shows two-dimensional planes
  of the four-dimensional basins of these points. For the
  $x \times y$ basins, $(z,w) = (0.1,0.0)$, and for the
  $z \times w$ basins, $(x,y) = (0.1,0.1)$. Black color is for
  periodic and red one is for chaotic solutions.}
  \label{fig8}
\end{figure}

\begin{figure}[htb]
  \centering
  \includegraphics*[width=1.00\columnwidth]{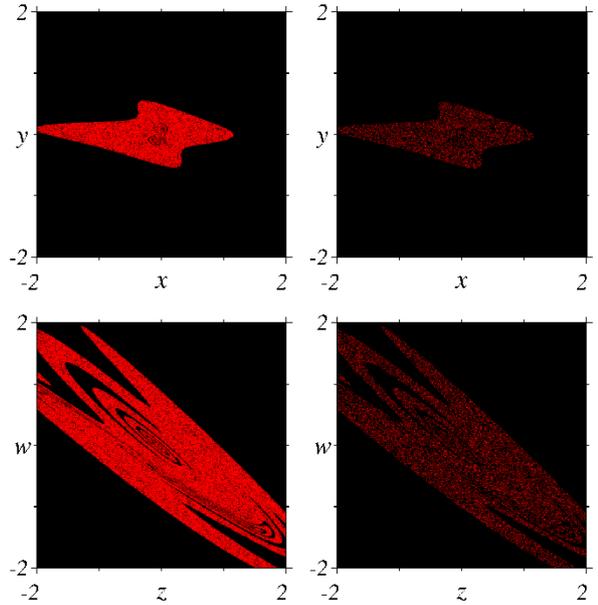}
  \caption{(Color online) Basins of attraction for the point B$_4$ in Fig.~\ref{fig3}(a),
  in the transient chaos region. Each column shows two-dimensional
  planes of the four-dimensional basins of this point. The left column
  is for a transient of $5.0 \times 10^5$, and the right column is for a
  transient of $4.5 \times 10^6$. For the $x \times y$ basins,
  $(z,w) = (0.1,0.0)$, and for the $z \times w$ basins,
  $(x,y) = (0.1,0.1)$. Black color is for periodic and red one
  is for chaotic solutions.}
  \label{fig9}
\end{figure}

To characterize the role of the initial conditions in the system~(\ref{chua4d}),
we investigate the basins of attraction for different sets of parameters. In
Figs.~\ref{fig8} and~\ref{fig9} we show some two-dimensional planes of the
four-dimensional basins of the points B$_1, \cdots, $B$_4$ in Fig.~\ref{fig3}(a).
The basins are obtained by evaluating the largest Lyapunov exponent, and
codifying its value with colors. Black color refers to null values and red one
refers to positive values, in which is considering a precision of $10^{-1}$
for positive values. Therefore, black color is the basin for
periodic attractors and red is the basin for chaotic attractors.
In Fig.~\ref{fig8}, the left column presents two planes of the
four-dimensional basin of attraction of the B$_1$ point in Fig.~\ref{fig3}(a).
This point is the focus of the spiral structure shown in Fig.~\ref{fig4},
roughly estimated at $(b,d) = (0.773,0.625)$. The central column presents
the basins for the B$_2$ point in Fig.~\ref{fig3}(a), for the parameters
$(b,d) = (0.95,5.50)$, and the right column presents the basins for the B$_3$ point
in Fig.~\ref{fig3}(a), for the parameters $(b,d) = (0.89,0.80)$.
In Fig.~\ref{fig9} we show the basin of attraction of the B$_4$ point
in Fig.~\ref{fig3}(a), for the parameters $(b,d) = (1.115,9.57)$.
This point is in the transient chaos region,
then the left column in Fig.~\ref{fig9} is for a transient time equal
to $5.0 \times 10^5$, and the right column is for $4.5 \times 10^6$.
In both figures it is clear to see that for points closer to the
transient chaos region, the basins of attraction are riddled
(B$_3$ and B$_4$ points), {\it i.e.}, in the chaotic attractor
domain (red region) there exist many black points one referring
to the periodic attractor domain, and as the transient time increases
the number of black points increases dominating the basin of the
chaotic attractor.

\section{Conclusions}
\label{conc}

The dynamics of a four-dimensional Chua system with smooth nonlinearity
was studied by numerical continuation and numerical solution methods.
By numerical solution methods, the Lyapunov and isoperiodic diagrams,
for the largest Lyapunov exponent and for the periods, respectively,
were used to show how rich is the dynamics presented by the system~(\ref{chua4d}).
In addition, by numerical continuation method, Hopf, period-doubling
and saddle-node bifurcation curves were constructed to show the
bifurcation structures on two-dimensional parameter-spaces.
	
With the largest Lyapunov exponents, was possible to observe periodic
structures immersed in the chaotic regions, and for the parameter $d$,
that controls the dimension of the system~(\ref{chua4d}), we observed
the presence of a periodicity spiral, where periodic structures are
connected between them and they coil up around a focal point.
This periodicity spiral was observed in
three-dimensional continuous-time systems \cite{bar2,gal8,gal9,alb2,slip},
recently in a four-dimensional continuous-time system \cite{gal10}, and
here we observed in a four-dimensional Chua model.
The parameter $d$ plays another role
in system~(\ref{chua4d}) regarding transient chaos.
We showed that for $d$ near zero, the occurrence of transient chaos
is small compared with $d$ greater than zero.

By studying the dynamics in the isoperiodic diagrams we
observed two different routes to chaos, at left in the diagrams
by period-doubling from periodic region to chaotic one,
and at right in the diagrams by crisis from chaotic region to
periodic one.

A recent paper \cite{gal9} proposes a definition of
{\it shrimp}-shaped domain as follows: {\it Shrimps are formed by
a regular set of adjacent windows centered around the main pair
of intersecting superstable parabolic arc. A shrimp is a
doubly infinite mosaic of stability domains composed by an
innermost main domain plus all the adjacent stability domains
arising from two period-doubling cascades together with their
corresponding domains of chaos}. In our work the bifurcation
curves revealed the endoskeleton of {\it shrimp}-shaped domains
embedded in chaotic regions. The endoskeleton is formed by
two saddle-node curves with one cusp point, and two intersected
period-doubling curves. The innermost main domain formed by these
four curves has the lowest period of the {\it shrimp}-shaped domain.
The period-doubling bifurcation curves showed the existence of a
{\it shrimp}-like domain of period-$2$ (see Fig.~\ref{fig2}) connected
to a large period-$1$ region, and these curves delineate the
borders of the lowest period of this structure.

\section*{Acknowledgments}
The authors thank CAPES, FAPESC, and CNPq for financial support.


\end{document}